\documentclass{article}
\usepackage{spconf,amsmath,graphicx,hyperref}
\usepackage{booktabs}

\title{MMAC: A Massive Multi-dimensional Benchmark for Audio Captioning}
\name{
\parbox{\linewidth}{\centering
Weijie Wu\textsuperscript{1,2}, Junbo Li$^2$, Lin Li$^3$, Jun Fang$^{*2}$, Qingyang Hong$^{*1}$
}
\thanks{\quad $^{*}$ Corresponding author.}
}
\address{
  $^1$School of Informatics, Xiamen University, China\\
  $^2$DiDi Global Inc., Beijing, China\\
  $^3$School of Electronic Science and Engineering, Xiamen University, China\\
  \texttt{vjjjjjj@stu.xmu.edu.cn}
  }

\begin{document}
\ninept
\maketitle
\begin{abstract}

With the development of audio large language models (AudioLLMs), audio captioning needs to move from brief descriptions toward open-ended and fine-grained free-form descriptions. Existing evaluations often focus on generation quality or task performance, making it difficult to diagnose information coverage and description reliability. We propose MMAC, a \textbf{M}assive \textbf{M}ulti-dimensional benchmark for \textbf{A}udio \textbf{C}aptioning. MMAC contains 5,638 audio clips from more than 20 data sources, covering 6 capability categories and 15 evaluation dimensions. Given a model-generated caption, MMAC checks whether it mentions relevant information in the target dimension and whether the mentioned content is consistent with the reference label. We evaluate representative open-source and proprietary AudioLLMs. Results show clear differences across evaluation dimensions, information coverage, and description reliability. We will release the MMAC benchmark and evaluation code.

\end{abstract}
\section{Introduction}

Audio captioning aims to convert audio signals into natural-language descriptions, so that models can summarize key information in audio with text. With the development of AudioLLMs~\cite{Qwen2.5-Omni,midashenglm,qwen3.5-omni,step-audio2}, captions are moving from brief descriptions to more open-ended and fine-grained audio understanding. In this setting, evaluation should not only consider whether the generated text is fluent or close to reference captions. It should also examine whether a model covers important audio information in free-form descriptions, and whether these descriptions are consistent with the audio. Therefore, detailed audio captioning requires evaluation beyond a single aggregate score, with attention to both information coverage and description reliability across different information dimensions and audio scenarios.

Existing benchmarks have advanced audio captioning and audio understanding from different perspectives. Some evaluations focus on the similarity between generated captions and reference descriptions, which measures the overall generation quality~\cite{mei2024wavcaps,kim2019audiocaps,drossos2020clotho}. Other evaluations convert audio understanding into more explicit test targets, and examine whether models understand certain types of audio information or specific application scenarios~\cite{MMAU,ma2026mmar,ma2026omnicaptioner}. These works provide important evidence for model comparison, but detailed audio captioning still requires further diagnostic evaluation. In particular, an aggregate score often cannot explain where model errors come from. A low score may result from omitted information, inaccurate descriptions of mentioned content, or mixed effects across different capability dimensions. This motivates a multi-dimensional diagnostic benchmark for detailed audio captioning, which can evaluate whether a model covers target information in the corresponding evaluation dimension and whether the mentioned content is consistent with the audio.

To this end, we propose MMAC, a multi-dimensional and fine-grained benchmark for audio captioning. MMAC decomposes detailed audio captioning into 6 capability categories and further divides them into 15 evaluation dimensions, covering different levels of audio information from spoken content and acoustic scenes to speaker attributes, speaking styles, temporal changes, and implicit meanings. During evaluation, all models receive the same open-ended caption prompt and generate natural-language descriptions. Unlike evaluations that only provide an aggregate score, MMAC does not require every caption to cover all dimensions. Instead, each test subset checks whether the model actively mentions the target information, and whether the mentioned content is consistent with the reference label. In this way, MMAC distinguishes omissions, incorrect descriptions, and correct descriptions, providing fine-grained diagnostic evidence for model comparison and error analysis. Our contributions are summarized as follows:

\begin{itemize}
    \item We introduce MMAC, a multi-dimensional and fine-grained benchmark for audio captioning. MMAC contains 5,638 audio clips from more than 20 data sources, covering 6 capability categories and 15 evaluation dimensions.

    \item We design a multi-dimensional diagnostic evaluation framework for free-form captions. Under a unified open-ended caption setting, the framework checks whether a model mentions target information in each evaluation dimension and further evaluates whether the corresponding description is accurate.

    \item We systematically evaluate representative open-source and proprietary AudioLLMs from the perspectives of Coverage, Precision, and Accuracy. The results reveal differences across evaluation dimensions, information coverage, and description reliability, and provide guidance for future audio captioning model development.
\end{itemize}

\section{MMAC}
\label{sec:MMAC}

\begin{figure*}[t]
\centering
\includegraphics[width=0.8\textwidth,keepaspectratio]{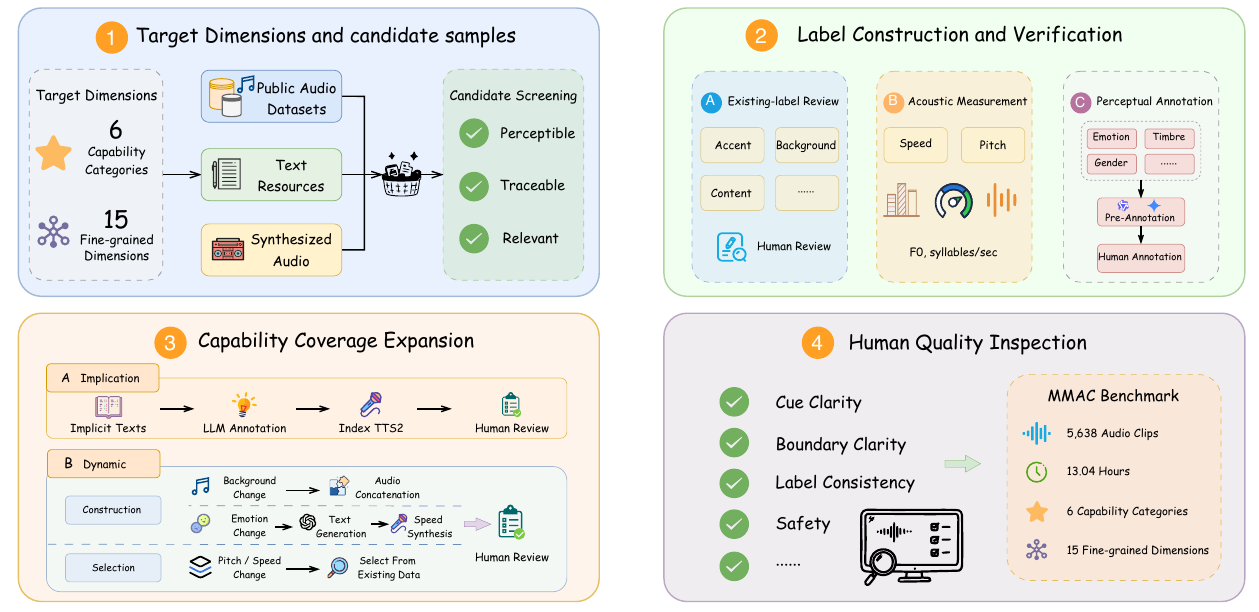}
\caption{Data construction and annotation pipeline of MMAC.}
\label{fig:annotation_pipeline}
\end{figure*}

\subsection{Overview}
\label{sec:Overview}

MMAC is a fine-grained evaluation benchmark for free-form audio captioning. Given an audio clip, it evaluates the natural-language caption generated by an audio language model across multiple dimensions. Instead of assigning only a holistic caption score, MMAC examines whether the caption covers target information and whether the corresponding description is consistent with the audio. This design provides a more detailed view of model behavior across different types of audio information.

Specifically, MMAC includes 5,638 audio clips from more than 20 data sources, covering 6 capability categories and 15 fine-grained dimensions. This design enables MMAC to reflect the overall captioning performance while revealing the strengths and weaknesses of different models.

\subsection{Benchmark Design}
\label{sec:benchmark_design}

MMAC is designed to provide a systematic, fine-grained, and diagnostic evaluation for free-form audio captioning. Traditional audio captioning evaluation usually relies on reference captions and n-gram based metrics such as BLEU and CIDEr, which mainly measure the overall closeness between generated and reference texts~\cite{af-next,secap}. In contrast, MMAC focuses on whether a model can actively cover specific audio information in free-form captions and describe it correctly. This design separates the coverage of target information from the correctness of the corresponding descriptions.

To define the evaluation scope of detailed audio captioning, MMAC organizes audio descriptions into 6 capability categories: \textit{content}, \textit{background}, \textit{persona}, \textit{paralinguistic}, \textit{dynamic}, and \textit{implication}. These categories correspond to spoken content, acoustic scenes and non-speech events, perceived speaker attributes, speech delivery, temporal changes, and implicit meanings beyond the literal content. Together, they cover audio information from perceptual details to higher-level semantic inference, and are further divided into 15 fine-grained evaluation dimensions. Table~\ref{tab:mmac_scale} reports the sample size and duration of each capability category.

We adopt a decoupled evaluation design and organize MMAC into multiple test subsets by capability category. Since audio samples usually do not contain all fine-grained dimensions at the same time, each test subset focuses on specific target dimensions to keep the evaluation objective clear and the labels reliable. All test subsets use the same open-ended caption prompt, and models always generate natural-language descriptions. Target dimensions are only used during evaluation to check whether the caption covers and correctly describes the relevant information. This design preserves the generation format of free-form captioning, while avoiding reformulating the evaluation as separate attribute question answering or classification tasks. It also allows MMAC to distinguish whether the model omits, incorrectly mentions, or correctly describes the target information. This distinction supports separate evaluation of information coverage and description correctness.

\begin{figure}[t]
    \centering
    \includegraphics[width=\linewidth]{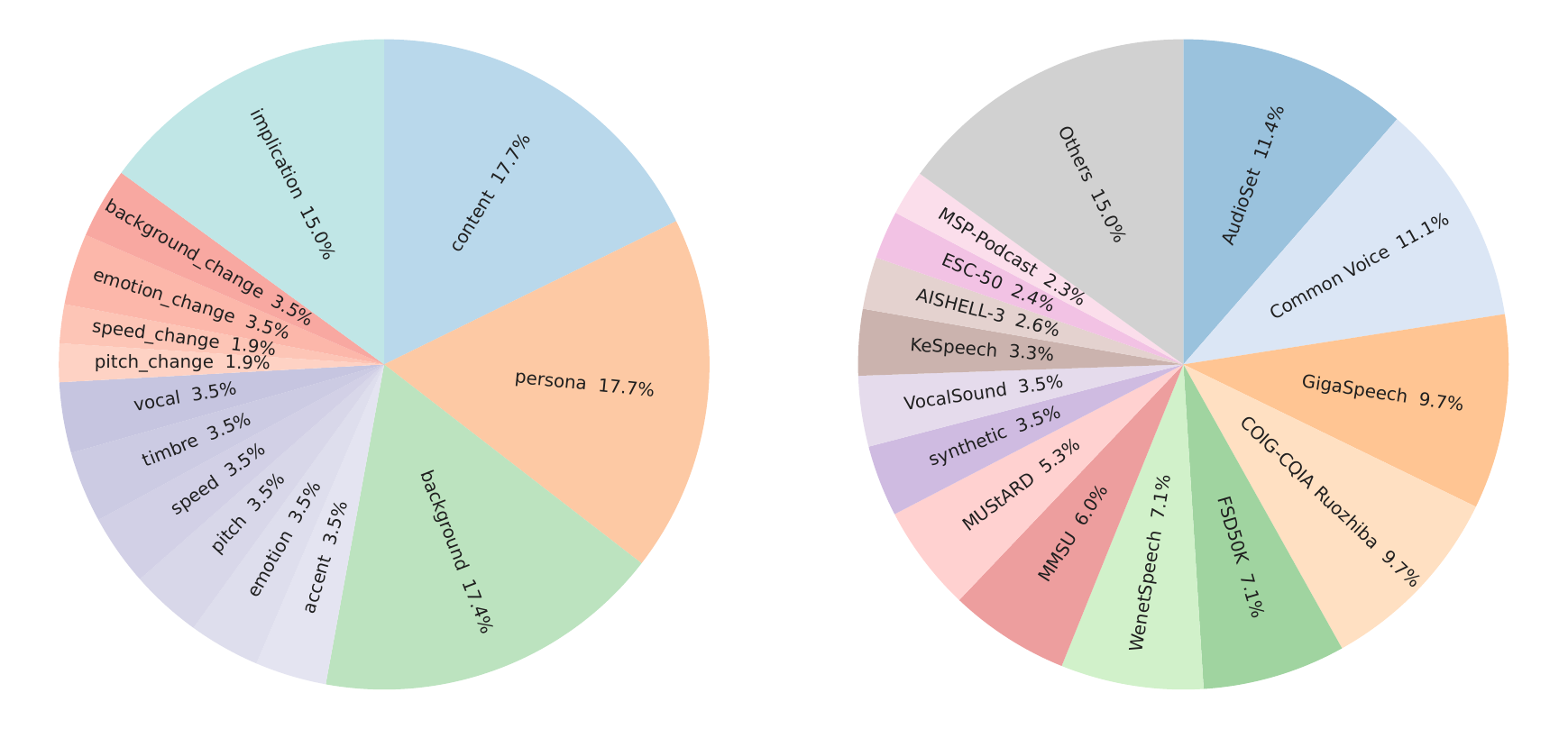}
    \caption{MMAC data statistics. Left: sample distribution across fine-grained dimensions. Right: sample distribution across data sources.}
    \label{fig:data_statistics}
\end{figure}

\begin{table}[t]
\centering
\caption{Statistics of MMAC by capability category. The overall duration is computed from unrounded values.}
\label{tab:mmac_scale}
\small
\begin{tabular}{l c c c}
\toprule
\textbf{Capability} & \textbf{\#Dims.} & \textbf{\#Samples} & \textbf{Duration} \\
\midrule
Content 
& 1 & 1000 & 2.27h \\

Background 
& 1 & 981 & 2.37h \\

Persona 
& 2 & 1000 & 1.88h \\

Paralinguistic 
& 6 & 1198 & 1.89h \\

Dynamic 
& 4 & 615 & 3.23h \\

Implication 
& 1 & 844 & 1.39h \\
\midrule
\textbf{Overall} 
& \textbf{15} & \textbf{5638} & \textbf{13.04h} \\
\bottomrule
\end{tabular}
\end{table}

\subsection{Data Collection}
\label{sec:data_collection}

MMAC is constructed around target dimensions rather than by simply merging existing datasets. We first collect candidate samples from public audio datasets, text resources, and synthesized audio, and then determine whether each sample can support the evaluation of a specific fine-grained dimension~\cite{MMSU,spark-tts, speechocean762, childmandarin, seniortalk}. A sample is retained according to three criteria: the target cue must be clearly perceptible in the audio, the label must have a traceable evidence source, and the key cue in the sample should be directly related to the target dimension. These criteria reduce attribution ambiguity caused by mixed capability cues and make subsequent error analysis more reliable. Following this principle, MMAC organizes data from different sources into 6 capability categories and 15 fine-grained dimensions, as illustrated in Fig.~\ref{fig:annotation_pipeline}.

During label construction, we retain reliable original labels whenever possible, and use additional review or measurement to make them suitable for evaluation at the dimension level. The content, background, and accent dimensions mainly rely on original transcriptions~\cite{gigaspeech,wenetspeech}, sound event labels~\cite{audioset,esc50,fonseca2021fsd50k}, and accent labels~\cite{kespeech}. Samples that may contain weak label noise are manually reviewed to remove cases where the target sound is unclear, the label is inconsistent with the audio, or the sample is difficult to judge. Pitch and speed are labeled using reproducible acoustic measurements, based on F0 and the number of syllables per second, respectively. Dimensions such as speaker attributes, emotion, and timbre require more auditory judgment, where original labels often suffer from differences in granularity, incomplete coverage, or unclear boundaries. We use Gemini 3.1 Pro\footnote{\url{https://gemini.google.com/}} and Qwen3-Omni~\cite{qwen3-omni} to verify candidate labels, and trained annotators further confirm, supplement, and revise them so that label descriptions match audible cues and label granularity is consistent across data sources.

Existing audio resources do not sufficiently cover all target dimensions, so we construct additional samples to complete the evaluation space. The implication dimension focuses on meanings behind speech. We select texts with implicit meanings, such as sarcasm and puns, from open-source text corpora~\cite{coig,mustard}, use Gemini 3.1 Pro to generate candidate explanations, and conduct human review to remove unsafe or semantically unclear samples. Inaccurate explanations are also corrected before the retained texts are synthesized into speech with Index TTS2. The dynamic dimension focuses on perceptible changes within an audio clip. To avoid relying only on naturally occurring changes, we combine audio composition, controlled text generation, and speech synthesis to obtain samples with clear change processes, covering changes in background, emotion, speed, and pitch. For emotion change, we prepare 200 topics for each of Chinese and English, and randomly pair each topic with two emotions. ChatGPT\footnote{\url{https://chatgpt.com/}} then generates texts with an emotional transition between the two segments, which are synthesized with IndexTTS2~\cite{indextts2}. Samples involving speed and pitch changes are selected from speech data containing the corresponding variations, and are further checked to ensure that the change process is clear and the boundary is identifiable. Before entering the final benchmark, all samples undergo human quality inspection. Samples are removed if the target cue is unclear, the change boundary is difficult to identify, or the label is inconsistent with the audio. Finally, MMAC contains 5,638 audio clips covering 6 capability categories and 15 fine-grained dimensions, with a total duration of 13.04 hours. Fig.~\ref{fig:data_statistics} summarizes the data composition and source distribution, while Table~\ref{tab:mmac_scale} reports the sample size and duration of each capability category.

\section{Experiments}
\label{sec:experiments}

\subsection{Experimental Setup}
\label{sec:experimental_setup}

We evaluate representative open-source and proprietary AudioLLMs on MMAC, including Qwen, Gemini, and other recent audio language models. All models receive the prompt \textit{``Describe this audio in detail.''} and generate free-form captions, except Qwen3-Omni-Captioner, whose inference interface does not support custom text prompts. We use the default generation parameters of each model and do not otherwise apply model-specific prompt tuning or decoding adjustments. Except for proprietary API models, all local inference is conducted on 8 NVIDIA A100 80GB GPUs. The generated captions are evaluated by Qwen3.6-27B\footnote{\url{https://huggingface.co/Qwen/Qwen3.6-27B}} under the same scoring rules. Fine-grained dimension scores are first averaged within each capability category, and the resulting scores of the 6 capability categories are then equally averaged to obtain the aggregate score.

\begin{figure*}[t]
\centering
\includegraphics[width=0.85\textwidth]{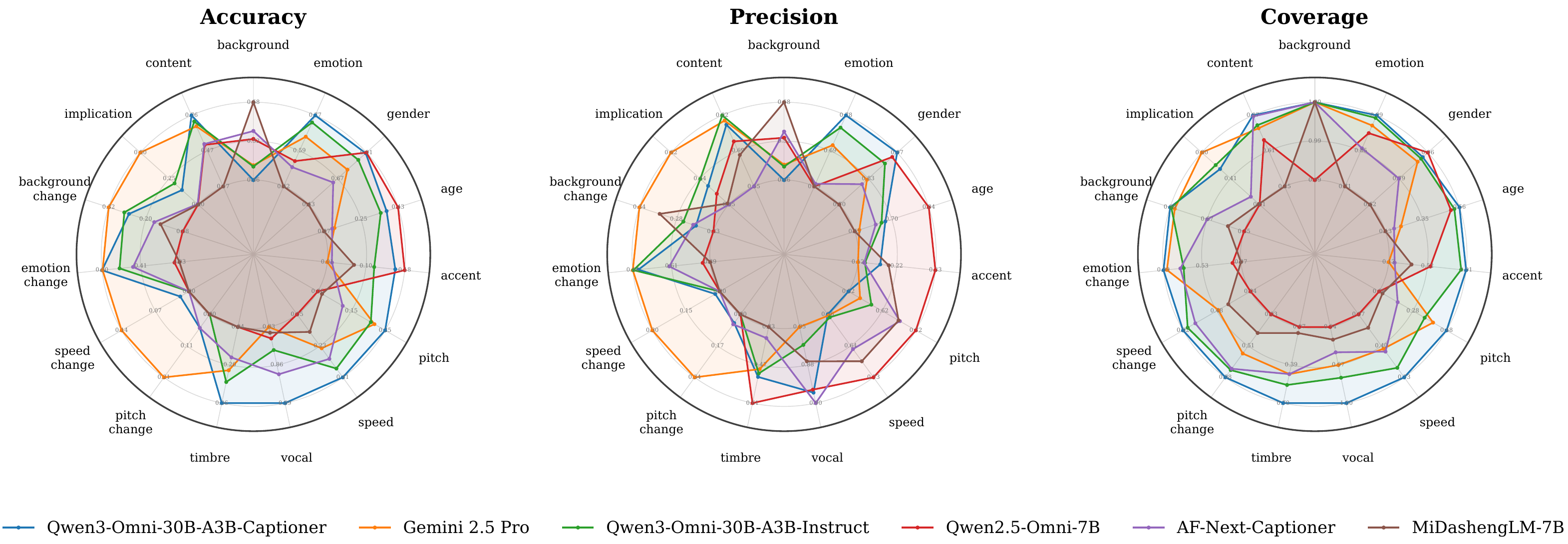}
\caption{Dimension-level performance of evaluated models on MMAC.}
\label{fig:radar_subdims}
\end{figure*}

\begin{table}[t]
\centering
\caption{Main results on MMAC. Scores are reported as percentages and computed by category-macro averaging.}
\label{tab:main_results}
\small
\setlength{\tabcolsep}{3.5pt}
\begin{tabular}{lccc}
\toprule
\textbf{Model} & \textbf{Accuracy} & \textbf{Precision} & \textbf{Coverage} \\
\midrule
Gemini 2.5 Pro & \textbf{46.85} & \textbf{59.39} & 75.31 \\
Qwen3-Omni-Captioner & 45.62 & 52.40 & \textbf{84.15} \\
Qwen3-Omni-Instruct & 42.96 & 54.22 & 78.90 \\
Gemini 2.5 Flash & 38.39 & 50.90 & 73.17 \\
Qwen2.5-Omni-7B & 33.60 & 50.92 & 51.22 \\
AF-Next-Captioner & 32.36 & 43.54 & 64.23 \\
Gemini 3.5 Flash & 26.18 & 50.67 & 50.49 \\
MiDashengLM-7B & 21.85 & 43.52 & 39.47 \\
\bottomrule
\end{tabular}
\end{table}

\subsection{Evaluation Metrics}
\label{sec:evaluation_metrics}

MMAC evaluates free-form descriptions generated under an open-ended caption prompt. Each sample is scored only on the target dimensions specified by its subset, rather than requiring every caption to cover all evaluation dimensions. Depending on the label type of each dimension, the judgment is mapped to either a binary or graded score, and all scores are normalized to $[0,1]$ before aggregation.

Based on this protocol, we report Coverage, Precision, and Accuracy. Coverage denotes the proportion of samples where the model mentions the target dimension. Precision denotes the average score over the mentioned samples. Accuracy denotes the average score over all valid samples, with omitted samples assigned a score of 0. The three metrics reflect information coverage, the reliability of mentioned descriptions, and the joint effect of coverage and correctness in captioning. Fine-grained dimensions are first averaged within each capability category. The resulting category scores are then equally averaged to obtain the aggregate score.

\subsection{Results}
\label{sec:results}

\begin{figure}[t]
\centering
\includegraphics[width=\linewidth]{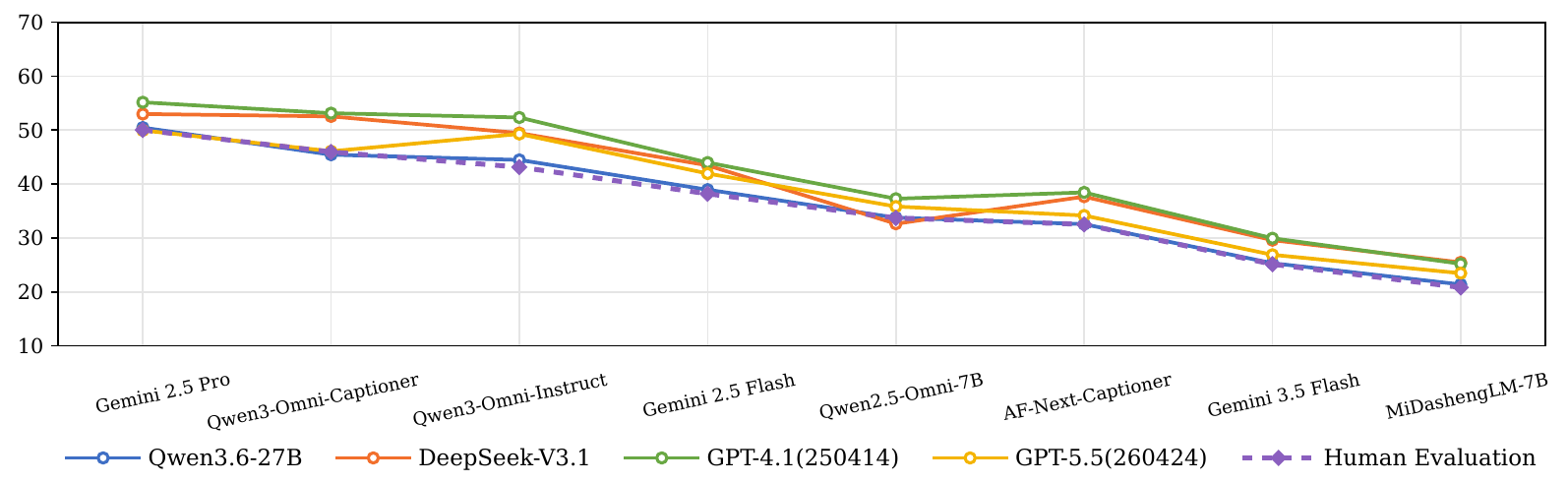}
\caption{Accuracy under different LLM judges and human evaluation on a 10\% subset stratified by fine-grained dimension.}
\label{fig:llm_judge_strict}
\end{figure}

Table~\ref{tab:main_results} reports the main results on MMAC. Gemini 2.5 Pro achieves the highest Accuracy and Precision, while Qwen3-Omni-Captioner obtains the highest Coverage. Gemini 3.5 Flash has competitive Precision but low Coverage, which limits its Accuracy. Together with Fig.~\ref{fig:radar_subdims}, these results reveal clear differences across fine-grained dimensions beyond the aggregate scores.

To further assess the stability of automatic evaluation, we perform stratified sampling within each fine-grained dimension and select 10\% of the samples. The outputs of all eight baseline models on the sampled data are evaluated using four LLM judges and human annotators. The human evaluation is conducted by three trained annotators. After annotation, a fourth annotator randomly audits 10\% of the human judgments, and the agreement between the original judgments and the audit exceeds 95\%. As shown in Fig.~\ref{fig:llm_judge_strict}, the Accuracy scores obtained from the four LLM judges and human evaluation follow similar trends and produce broadly consistent model rankings. The five rankings yield a Kendall's coefficient~\cite{Kendall} of concordance of $W=0.981$, indicating that the relative model performance identified by MMAC remains stable across different judges and is broadly consistent with human assessment despite differences in absolute scoring scales.

\subsection{Analysis}
\label{sec:analysis}

Sec.~\ref{sec:results} shows that models with comparable Accuracy can still differ substantially across MMAC dimensions. Gemini 2.5 Pro performs better on implication and dynamic, indicating stronger performance in inferring implicit meanings and describing temporal changes. In contrast, Qwen3-Omni-Captioner provides broader Coverage on descriptive dimensions such as content and paralinguistic information. This suggests that a single aggregate score cannot fully capture how models differ across dimensions.

\begin{figure}[t]
\centering
\includegraphics[width=\linewidth]{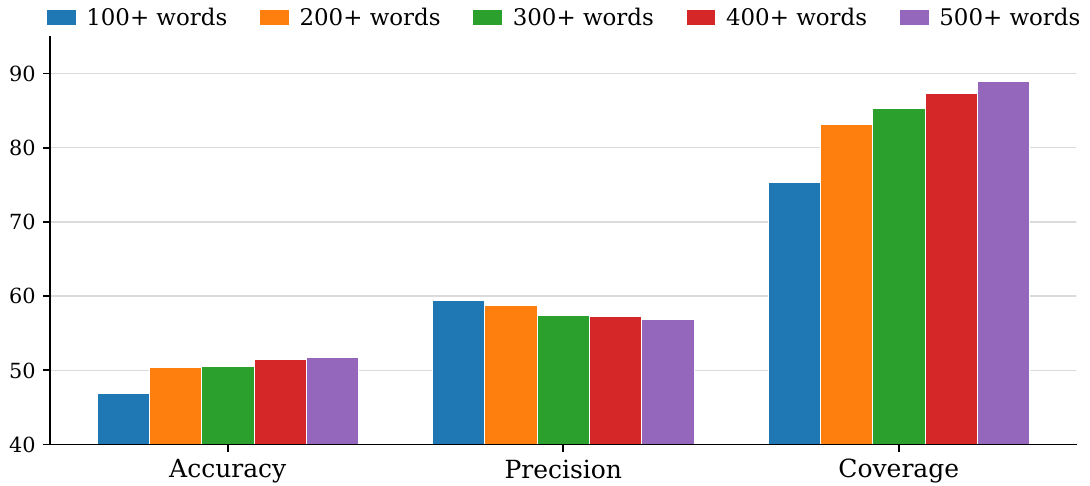}
\caption{Effect of caption length on MMAC evaluation.}
\label{fig:length_control}
\end{figure}

The gap between coverage and reliability is also reflected in model training and output length. Comparing the two Qwen3-Omni models, supervised fine-tuning for captioning substantially improves Coverage, but does not lead to consistent Precision gains. Implication is the only dimension where both Coverage and Accuracy decrease, suggesting that a stronger tendency to describe more information does not necessarily improve implicit meaning inference. We further control the output length of Gemini 2.5 Pro by adding target word-count constraints to the original prompt. As shown in Fig.~\ref{fig:length_control}, longer captions cover more target information, but Precision decreases as length increases. These results show that detailed audio captioning should be evaluated in terms of both information coverage and description reliability.

\section{Conclusion and future work}
\label{sec:conclusion}

In this paper, we presented MMAC, a multi-dimensional benchmark for audio captioning. MMAC comprises 6 capability categories and 15 fine-grained dimensions and evaluates the coverage and correctness of target information in free-form captions. Results reveal clear differences among AudioLLMs in information coverage and description reliability. Analyses of captioning-oriented supervised fine-tuning and output length further show that broader coverage does not necessarily lead to more reliable descriptions. MMAC provides a fine-grained basis for model comparison, error analysis, and future captioning model development. One limitation is that our human evaluation used pre-annotations generated by Qwen3.6-27B, which may have introduced anchoring bias. Therefore, the close agreement between the scores produced by Qwen3.6-27B and the human ratings should not be interpreted as evidence that this judge is superior to the alternatives. Future work will adopt independent annotations, extend MMAC's coverage of languages, scenarios, and dimensions, and introduce timestamp-based evaluation for more precise temporal localization.

\bibliographystyle{IEEEbib}
\bibliography{strings,refs}

\end{document}